\documentclass[intlimits,twoside,a4paper]{article}
\usepackage{amsmath,amssymb}
\usepackage{graphicx}
\usepackage{subfig}
\usepackage[T2A]{fontenc}
\usepackage[cp1251]{inputenc}

\usepackage{cmpj2}

\issue{2015}{18}{3}{33801}
\doinumber{10.5488/CMP.18.33801}

\title[Elastic properties of 5$d$ transition-metal carbides: An {\it ab initio} study]%
{Elastic properties of 5$d$ transition-metal carbides: \\An {\it ab initio} study}

\author[L. Mex, A. Aguayo, G. Murrieta]{L. Mex\refaddr{label1}, A. Aguayo\refaddr{label2},
G. Murrieta\refaddr{label2}}

\addresses{
\addr{label1}
Facultad de Ingenier\'{i}a, Universidad Autynoma de Yucat\'{a}n. Av. Industrias no Contaminantes por Perif\'{e}rico Norte Apdo. Postal 150 Cordemex
M\'{e}rida, Yucat\'{a}n, M\'{e}xico
\addr{label2} Facultad de Matem\'{a}ticas, Universidad Autynoma de Yucat\'{a}n. Perif\'{e}rico Norte, Tablaje 13615, C. P. 97110, M\'{e}rida, Yucat\'{a}n, M\'{e}xico
}

\date{Received December 9, 2014, in final form April 14, 2015}
\begin{document}

\maketitle

\begin{abstract}
We have systematically studied the mechanical stability of group V transition
metal carbides TMC$_2$ (TM${}={}$Hf, Ta, W, Re, Os, Ir, Pt, and Au) in the pyrite and
fluorite phase, by calculating their elastic constants within the density
functional theory scheme. It was found that all but ReC$_2$ and OsC$_2$ are
stable in pyrite phase. On the other hand, all metal carbides studied were
unstable in the fluorite phase.

\keywords first-principles calculation, elastic constants, hard material, transition metals

\pacs 81.05.Je, 81.05.Zx, 71.15.Mb, 71.20.Be

\end{abstract}

\section{Introduction}

The elastic stability criterion has been used to establish the
possible stability or metastability of a crystallographic phase in the
solid state. It has been shown that through the first-principles
calculation it is possible to obtain the elastic properties and to
describe the strength of the bond between neighboring atoms. Since
the adequacy of the theoretical results made it possible to adequately
predict the formation of new crystal structures,
it became possible to synthesize some of the predicted structures.

Intensive theoretical and experimental efforts have been focused on the
possibility of finding new low compressibility materials with hardness
comparable with diamond~\cite{Jhi1999}. Superhard materials are of
primary importance in modern science and technology due to
their
numerous applications, starting from cutting and polishing tools up to wear resistant
coatings. In this search, special interest has been taken in the metal
carbides and nitrides. The introduction of smaller atoms such as nitrogen or carbon into interstitial sites in closely packed transition metal lattices changes their chemical and physical properties
with respect to the metal. Transition metal carbides and nitrides have been considerably investigated due to their unique chemical and physical properties, such as high thermal conductivity,
high melting point, chemical inertness, high stiffness, high hardness, and metallic electrical
conductivity~\cite{Sahnoun2005,Friedrich2011,Friedrich2010,Srivastava2012,Srivastava2011,Chauhan2013}. Transition metal carbides or nitrides had been little studied due to
the difficulty of obtaining the crystalline samples. However, a series of
works such as PtN, PtN$_2$, IrN$_2$, and PtC~\cite{ono2005}
triggered a rapid advance in the production of new carbide and nitride
materials. They can have different morphologies, atomic structures and
substantially differ from the corresponding crystalline phase relative to their physicochemical
properties. To date, a wide group of nanocarbides of the $d$ metals
such as molecular clusters, nanocrystals, nanospheres, nanowires,
nanotubes, etc.,
have been synthesized. Several of these new systems formed
by transition metals and nitrogen or carbon
have found no consensus on
their crystal structure. Among all these materials,   the group of platinoid nitrides
and carbides attracted  a
particular interest due to
their technological potential. Crowhurst et
al.~\cite{science311} in order to explain the high bulk modulus of
platinum nitrides studied the Pt--N system in the stoichiometry
PtN$_2$. They analyzed the composite in two different structures,
fluorite and pyrite, and found that the most stable phase is the pyrite
phase, with an internal parameter $u$ = 0.415.

To explore
the possible existence of 5$d$ transition-metal carbides
with C/M = 2 stoichiometry in the pyrite or fluorite phases, using
density functional theory, we calculated the elastic constants in both
cubic phases and
analyzed
the performance of elastic stability criteria.

\section{Methods}    

The first-principles calculations were performed based on the density
functional theory (DFT). The Kohn-Sham total energies were
self-consistently calculated using the linearized augmented plane wave method
(FP-LAPW) with local orbital extensions~\cite{Singh2006}, as
implemented in the WIEN2k~\cite{Blaha2013,Schwarz2003} code, where the
core states are treated fully relativistically, and the semicore and
valence states are computed in a scalar relativistic approximation.
The exchange-correlation terms were considered in the
Perdew-Burke-Ernzerhof form of the generalized gradient approximation
(GGA)~\cite{ggaPBE96}. We have chosen the muffin-tin radii
(R$_{\text{MT}}$) of 2.0 a.u. for the transition metals and 1.2 a.u.
for the carbon atoms. The self-consistent calculations were done with
an LAPW basis set defined by the cutoff
$R_{\text{MT}}K_{\text{MAX}}$=8.0. Inside the atomic spheres, the
potential and charge densities are expanded in crystal harmonics up to
$L=10$. Convergence was assumed when the energy difference between the
input and output charge densities was less than $1\times 10^{-5}$ Ry.
The calculations were carried out with a sufficiently large number of
$\mathbf{k}$ points in the first Brillouin zone (BZ). We used a $13\times
13\times 13$ $\mathbf{k}$-point mesh, yielding a different number of $\mathbf{k}$ points in
the irreducible wedge of the BZ depending on the structure: 256 for
the fluorite, and 176 for the pyrite phase.

We evaluated the structure of TMC$_2$ (TM=Hf, Ta, W, Re, Os, Ir, Pt,
and Au) in the pyrite and fluorite phase. Pyrite (FeS$_2$ structure type)
has a cubic crystal structure with space group Pa$\overline{3}$ (205), the
transition metal atoms occupying Wyckoff site $4a$ (0,0,0), and the
carbon atoms are grouped as dimers around the fcc octahedral
interstitial sites oriented in the $\left<111\right>$ directions at
$8c$ ($u,u,u$). The transition metal in the pyrite structure  is fixed
by symmetry but the carbon atoms
have one free parameter ($u$)~\cite{pyrite}. Fluorite is a particular high
symmetry phase of pyrite. When the free parameter $u = 0.25$ at the 8c site, we
obtain the fluorite (CaF$_2$ structure type) structure with space group cF12 (225).

The calculated total energy as a function of volume was fitted to the
Birch--Murnaghan equation of state~\cite{birch}. From this process, the
equilibrium lattice constant ($a$) and the bulk modulos ($B$) were
obtained. All crystals in a cubic structure have only three independent
elastic constants, namely $C_{11}$, $C_{12}$, and $C_{44}$.
One can use a small strain and calculate the change of energy or stress
to obtain elastic constants $C_{ij}$. In the crystal structures analyzed in this work, an external strain $\delta$ from $-0.08$ to + $0.08$ was applied in the directions as
explained by G\"{u}emez et al.~\cite{alex2011}, associated with
deformations: isotropic, tetragonal and orthorhombic.

\begin{equation}
\epsilon_{\mathrm{iso}}=\left(
\begin{array}{ccc}
(1 + \delta)^{1/3} &  0  &  0 \\
 0  &  (1 + \delta)^{1/3} &  0 \\
 0  &  0  & (1 + \delta)^{1/3}
\end{array}
\right),\nonumber 
\end{equation}

\vspace{0.3cm}

\begin{equation}
 \epsilon_{\mathrm{tet}} = \left(
  \begin{array}{ccc}
  (1+\delta)^{-\frac{1}{3}} & 0 & 0 \\
  0 & (1+\delta)^{-\frac{1}{3}} & 0 \\
  0 & 0 & (1+\delta)^{\frac{2}{3}} \\
  \end{array}
  \right),
  \label{estabilidad}
  \end{equation}

\vspace{0.3cm}

\begin{equation}
\epsilon_{\mathrm{ort}} =  \left(
  \begin{array}{ccc}
  1 & \delta & 0 \\
  \delta & 1 & 0 \\
  0 & 0 & 1+\delta^2 \\
  \end{array}
  \right),\nonumber 
  \end{equation}
to distort the lattice vectors, $R'=(1+\epsilon)R$. The resulting
changes of energy are associated with elastic constants,
$$
\Delta E_{\mathrm{iso}}=\frac{V_0}{2}(C_{11}+2C_{12})\delta^2 =\frac{2V_0}{3}B\delta^2,
$$
$$
\Delta E_{\mathrm{tet}}=\frac{V_0}{3}(C_{11}-C_{12})\delta^2 \quad\text{and}
$$
$$
\Delta E_{\mathrm{ort}}=2V_0C_{44}\delta^2.
$$

In order to be mechanically stable, the elastic stiffness constants of a given
crystal should satisfy the generalized elastic stability
criteria~\cite{Jinghan1993}. The elastic stability criteria for a cubic
crystal at ambient conditions are,

\begin{equation}
\label{conditions}
C_{11}+2C_{12}>0,\quad C_{11}-C_{12}>0,\quad\text{and}\quad
C_{44}>0,
\end{equation}
\noindent
i.e., all the bulk moduli ($B$), shear ($C_{44}$), and tetragonal shear
[$C'=(C_{11}-C_{12})/2$] moduli are positive.
We also calculated the Young's modulus ($E$),
which provides a measure of stiffness and stability of the solids.
Another interesting elastic
property for any applications, particularly for their anisotropy, is the Zener
factor $A$. These quantities are calculated in terms of the
computed $C_{ij}$ using the following relations
\begin{equation}
\label{young}
E=\frac{9BG}{3B+G}\,,
\end{equation}
\begin{equation}
\label{zener}
A=\frac{C_{44}}{C'}=\frac{2C_{44}}{C_{11}-C_{12}}\,,
\end{equation}
where $G$ is the isotropic shear modulus. For a cubic material
with its
two shear constants, $C_{44}$ and $C'$, the value of $G$ should be between
these two constants. Therefore, G has a unique value if $C_{44}$ =
$C'$. This happens if the material is isotropic, that is, the Zener
factor is equal to one, as in the case of W. By assuming a homogeneous
strain on the compound, Voigth~\cite{voight1928lehrbuch}
established the upper limit of $G$ as
\begin{equation}
\label{shearmodulusvoigt}
G{\mathrm{V}}=\frac{1}{5}C'(2+3A).
\end{equation}

On the other hand, assuming a homogeneous stress,  as the lower bound Reuss~\cite{reuss1929calculation}
proposes
\begin{equation}
\label{shearmodulusreuss}
G{\mathrm{R}}=5C'\frac{A}{3+2A}\,.
\end{equation}

In this work, we take the
arithmetic average as proposed by Hill~\cite{Hill1952}
\begin{equation}
\label{shearmodulus}
G=\frac{1}{2}(G{\mathrm{V}} + G{\mathrm{R}}).
\end{equation}


\section{Results and discussion}

\begin{figure}[htb]
\begin{center}
\includegraphics[width=0.65\textwidth]{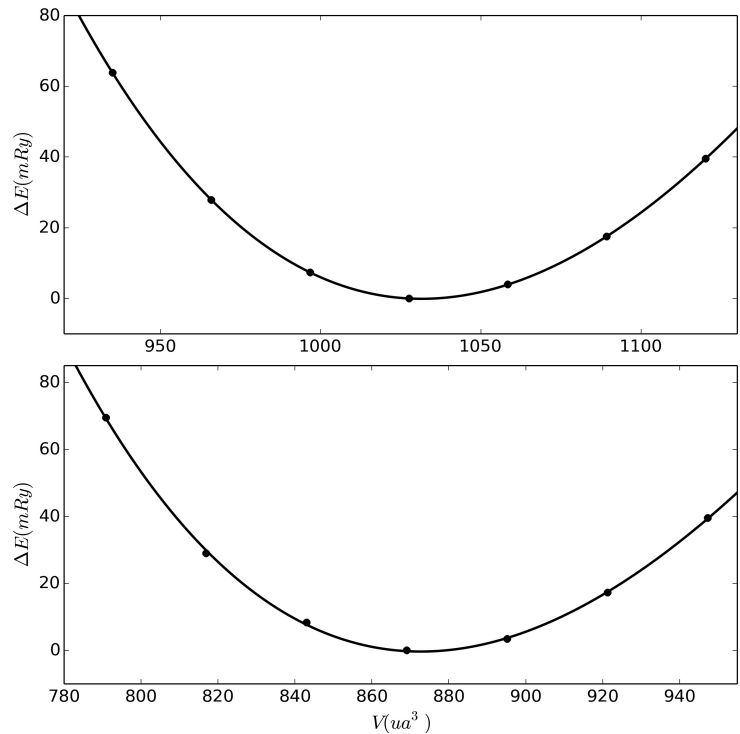}
\caption{Calculated total energies for HfC$_2$(top) and AuC$_2$(bottom) under isotropic deformation (circles). The energy of the pyrite phase at the equilibrium volume is the reference level, and is set to zero. The line corresponds to a fit of the Birch-Murnaghan equation of state to the calculated energy values (see text).}
\label{eosfitHfandAu}
\end{center}
\end{figure}

In figure~\ref{eosfitHfandAu} we show the calculated total energy of pyrite
HfC$_2$ and AuC$_2$ (pyrite-face) for seven values of the cell volume
(open circles). The calculated total energy as a function of volume was fitted
to the Birch-Murnaghan equation of state\cite{birch}. The fit is presented
in figure~\ref{eosfitHfandAu} (solid line), were the energy is given with
respect to the minimum energy of the pyrite structure.

\begin{figure}[!h]
\begin{center}
\includegraphics[width=0.65\textwidth]{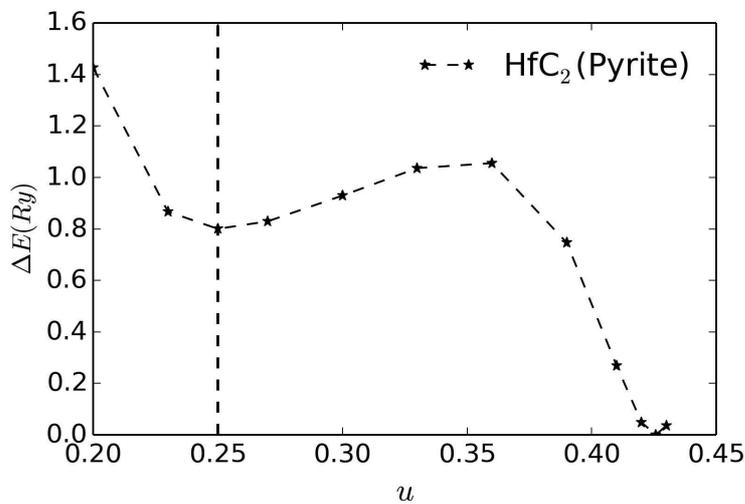}
\caption{Energy of HfC$_2$ as a function of the position of C atoms ($u$).
When $u=1/4$,
HfC$_2$(pyrite) reduces to HfC$_2$(fluorite). For all the  transition metal carbides
studied in this work, the second minimum was in $u = 0.43$ }
\label{HfEvsu}
\end{center}
\end{figure}

In order to find the value of the free parameter $u$ in the pyrite
phase, the DFT total energy and the forces acting on the atoms were
optimized. Figure~\ref{HfEvsu} shows the DFT total energy as a
function of the free parameter $u$ for HfC$_2$. In this curve,
we have got two local minima, one of them being observed at $u=0.25$, which
correspond to a possible metastable fluorite phase. At $u$ = 0.43, there
is another minimum
at much lower energy, which
corresponds
to pyrite phase. In all the 5$d$ transition metal carbides studied in
this work,
we have got
the same value for the free parameter $u$. The
difference in the energy between the two minimum phase suggests that
pyrite phase should be a more stable phase at zero pressure.

From the analysis of elastic stability, applying a standard set
of deformations
[equation~\eqref{estabilidad}], we find that
all compounds
in fluorite phase are unstable, particularly under the tetragonal deformation,
that is, the relation $C_{11} - C_{12}$ turned out to be negative.

The table~\ref{table} presents the results of our calculations for the
pyrite phase. Our results demonstrate that HfC$_2$, TaC$_2$, WC$_2$,
IrC$_2$, PtC$_2$ and AuC$_2$ are mechanically stable, while ReC$_2$
and OsC$_2$ violate the mechanical stability conditions. In both of
them, the shear modulus (C$_{44}$) was negative, so the shear modulus
$C_{44}$ is the main constraint on stability in those compounds in the
pyrite phase. As can be
seen, in general it holds for pyrite phase that
$B>C'>C_{44}>0$.
\begin{table}[*h]
\caption{DFT lattice constant $a$, zero pressure
elastic constants $c_{ij}$ (GPa), bulk modulus $B_0$ (GPa), shear
modulus $G$ (GPa), Zener factor $A$, and Young's modulus
(GPa), calculated in the present work for pyrite phase of period
VI transition metal carbides.}
\vspace{2ex}
\label{table}
\begin{center}
\begin{tabular}{|c|c|c|c|c|c|c|c|c|c|}
\hline
        &$a$(\AA)&C$_{11}$ &C$_{12}$ &C$_{44}$ &$B_0$ &$G$&$A$ &$E$ \\\hline\hline
HfC$_2$ &5.34  &275            &127           &53          &176     &61 & 0.72 &158\\
TaC$_2$ &5.12  &380            &168           &71          &238     &83 & 0.67 &217\\
WC$_2$  &5.03  &432            &183           &30          &266     &55 & 0.24 &152\\
ReC$_2$&4.98   &Unstable &Unstable  &Unstable &282         &--      & --& --     \\
OsC$_2$&4.96   &Unstable &Unstable  &Unstable &286         &--      & --& --     \\
IrC$_2$&4.95   &569            &133           &69          &279     &112 & 0.32 &289\\
PtC$_2$&4.98   &603            &101           &88          &268     &136 & 0.35 &342\\
AuC$_2$&5.06   &445            &102           &116         &217     &135 & 0.68 &326\\
\hline
\end{tabular}
\end{center}
\end{table}

\begin{figure}[h]
\includegraphics[width=0.48\textwidth]{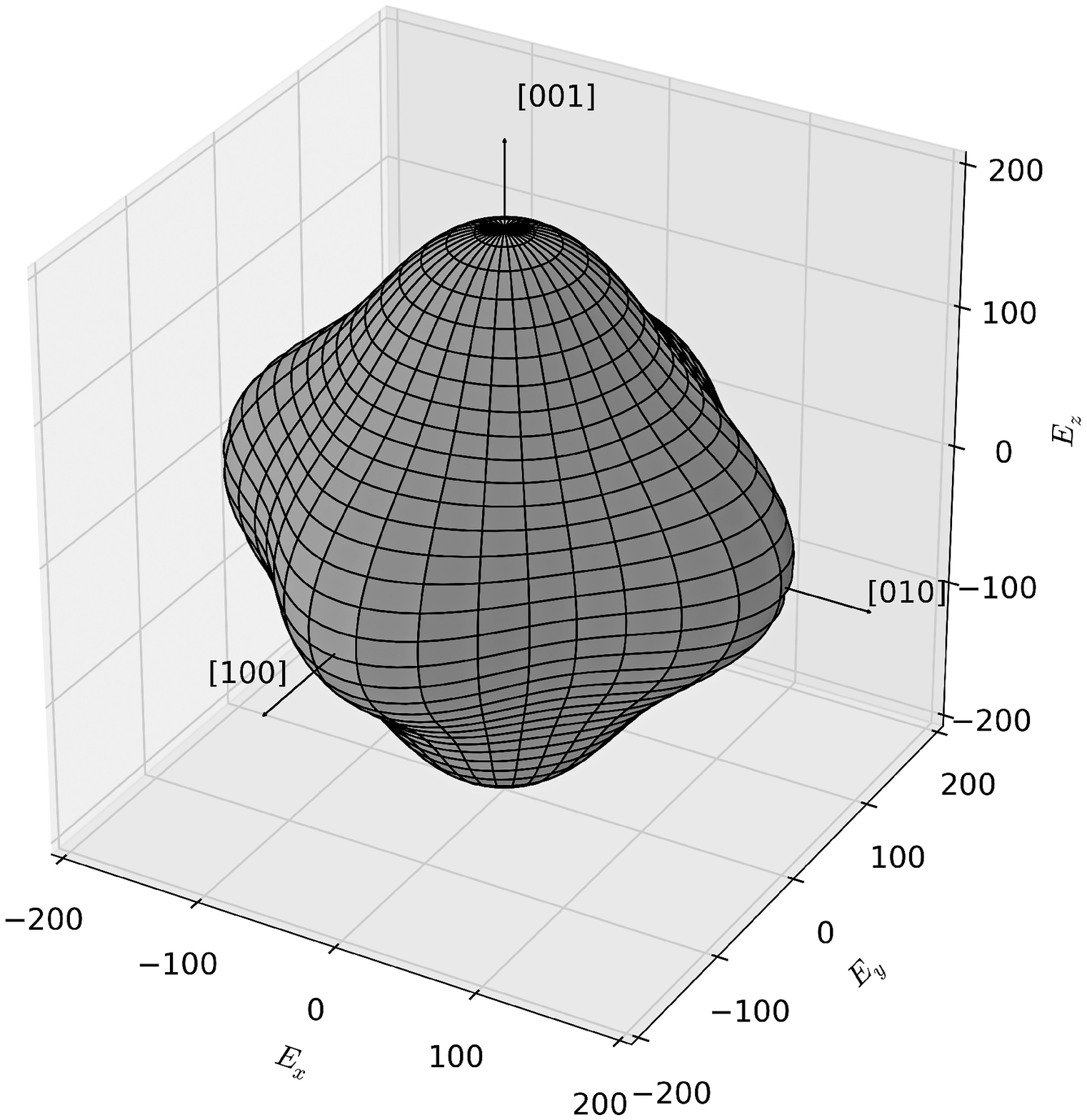}%
\hfill%
\includegraphics[width=0.48\textwidth]{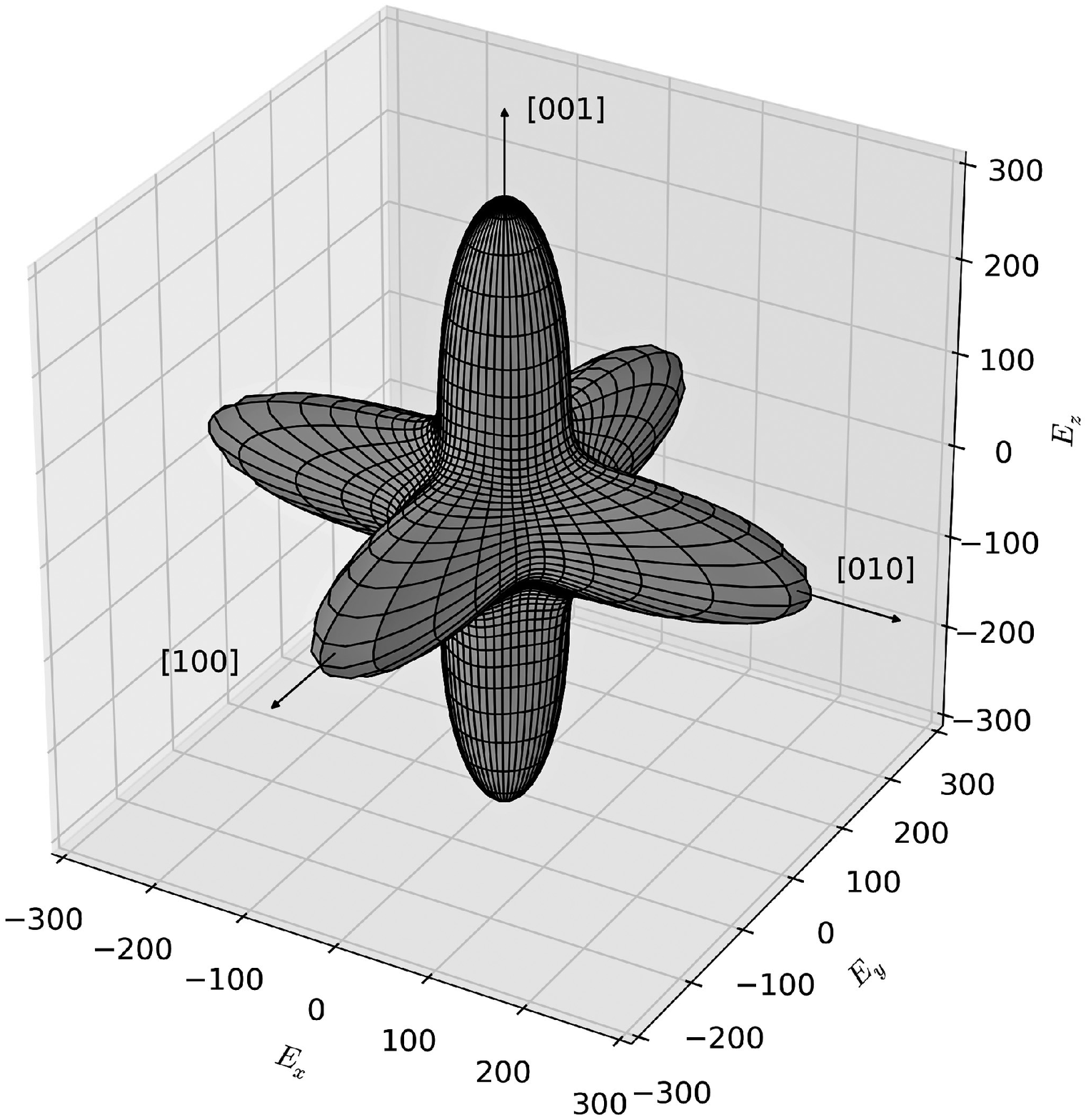}%
\\%
\parbox[t]{0.48\textwidth}{%
\centerline{(a)}%
}%
\hfill%
\parbox[t]{0.48\textwidth}{%
\centerline{(b)}%
}%
\caption{3D directional dependence of the Young's modulus for
(a) AuC$_2$  ($A$ = 0.68) and (b) WC$_2$ ($A$ = 0.24).}
\label{YoungModulosSurAuC2}
\end{figure}

The Zener anisotropy factor $A$ is a measure of the degree of elastic
anisotropy in solids. $A$ will take the value of 1 for a completely
isotropic material.
A
value of $A$ smaller or greater than unity
shows the degree of elastic anisotropy. The calculated Zener
anisotropy for pyrite structure (see table~\ref{table}) implies
that all compounds are elastically anisotropic ($A<$1). In order to better visualize the anisotropy of these compounds, we show a
three-dimensional (3D) representation of Young's modulus. For cubic
crystals, the directional dependence of the Young's modulus in 3D
representation can be given by
\begin{equation} 
\frac{1}{E} = S_{11} - 2\left(S_{11}
- S_{12} - \frac{1}{4}S_{44}\right)\left(l_1^2l_2^2+l_2^2l_3^2+l_3^2l_1^2\right),
\end{equation}
where $S_{ij}$ are the elastic compliance constants, and
$l_1$, $l_2$ and $l_3$ are the directional cosines to the $x-$, $y-$
and $z-$axes, respectively.
In the $\left< 100 \right>$ directions, the
second term is zero, and for $C_{44}/C'<1$, a maximum
in $\left< 100 \right>$ directions.
In all the compounds, the Zener's coefficient was
less than one, so that $\left<111 \right>$ directions are soft and
$\left< 100 \right>$ are hard. Three-dimensional representation of Young's
modulus is shown in figure~\ref{YoungModulosSurAuC2}. It can be seen that the elastic
anisotropy increases in the direction $\left< 100 \right>$ as Zener's
coefficient decreases.

The bulk moduli follow
the parabolic behavior, and increases from HfC$_2$ to reach the maximum on IrC$_2$. This behavior is similar to that present in the
corresponding 5$d$ transition metals. However, not all compounds,
the bulk modulus  increases relative to that of the pure elements, as in the
case of nitrides. The values of the bulk
modulus increase in: HfC$_2$ (from 109 GPa to 176 Gpa), TaC$_2$
(from 200 GPa to 238 GPa),  PtC$_2$ (from 230 to 268), and AuC$_2$ (from 173
to 217). On the other hand, the other two stable compounds in the pyrite phase
show a decrease in the bulk modulus
compared to
that of pure transition metals: WC$_2$
from 323 GPa to 266 GPa, and  IrC$_2$ from 355 GPa to 279 GPa.
This same trend is followed by the Young's modulus with an increasing value
in the dicarbides with Hf, Ta, Pt and Au.

\begin{figure}[*h] \begin{center}
\includegraphics[width=0.65\textwidth]{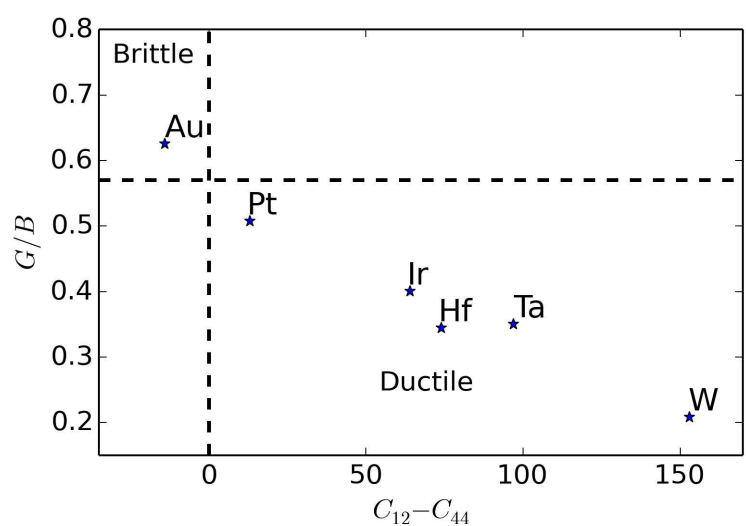}
\caption{Map of brittleness and ductility trends of HfC$_2$, TaC$_2$, WC$_2$, IrC$_2$,
PtC$_2$, and AuC$_2$. In the figure, only the transition metal simbol is shown.} \label{mapbritteandductile}
\end{center}
\end{figure}

The generally accepted rule is to
associate hard compounds with high bulk and shear modulus values
\cite{Bocquillon2001}. However, bulk modulus is not the only mechanical
quantity that determines the utility of a material for hard coatings.
Another important material
property to be also considered is
toughness, which is influenced by the degree of plastic deformation
(ductility) of the material under mechanical loading.
To analyze the ductility of the compounds, we use the Pettifor's
criterion~\cite{pettifor1992,Sangiovanni2011}, which states that, for
metallic non-directional bonding compounds, the Cauchy pressure
($C_{12}-C_{44}$) value is typically positive. This region corresponds
to a ductile behavior of a material. The other criterion we used was the
Pugh's modulus ratio G/B, If G/B $>$ 0.57\cite{pugh1954},
and the materials behave in a brittle manner. In
figure~\ref{mapbritteandductile}, we show the Pugh and Pettifor
criteria to map the ductility and brittle behavior for pyrite phase of
period VI transition metal carbides. According to
their formulation, the
only compound which is not within the ductile region of the map is
AuC$_2$.

\section{Conclusion}

In summary, we have studied metal carbides TMC$_2$ in the fluorite and
pyrite structures using first-principles calculations. With all the
calculations, we conclude that
all metal carbides studied with fluorite structure
are mechanically unstable. On the other hand, the
pyrite phase for ReC$_2$ and OsC$_2$ does not meet the stability
criteria as well. The bulk and Young's moduli value increases to the
value of the pure elements in HfC$_2$, TaC$_2$, PtC$_2$,
and Au$_2$, but
decreases in WC$_2$ and IrC$_2$.
According to the criterion of brittleness (ductility), all compounds
except AuC$_2$  exhibit a ductility behavior.

\section*{Acknowledgements}

The authors would like to thank Carlos Brito
for helpful comments. This work was supported by Facultad
de Mat\'amaticas-UADY under Grant no. FMAT--2012--0007 and CONACyT under Grant
no. 025794.

\ukrainianpart \label{ukr}
\title{Еластичні властивості карбідів  5$d$ перехідних металів: \\ {\it ab initio} дослідження}

\author{Л. Мекс\refaddr{label1}, А. Агуайо\refaddr{label2}, Г. Мурріета\refaddr{label2}}

\addresses{
\addr{label1} Інженерний факультет, Автономний університет Юкатану, Юкатан, Мексика
\addr{label2} Факультет математики, Автономний університет Юкатану, Юкатан, Мексика
}

\makeukrtitle
\begin{abstract}
\tolerance=3000%
Проведено систематичні дослідження менханічної стійкості карбідів перехідних металів п'ятої гру\-пи
TMC$_2$ (TM${}={}$Hf, Ta, W, Re, Os, Ir, Pt, і Au) у піритовій і флюоритовій фазах
шляхом обчислення їх еластичних сталих в рамках  теорії функціоналу густини.
Встановлено, що всі карбіди металів, за винятком
ReC$_2$ і OsC$_2$, є стійкими у піритовій фазі. З іншого боку, всі метали карбідів, що вивчалися, виявилися нестійкими у
флюоритовій фазі.

\keywords першопринципні обчислення, еластичні стали, твердий матеріал, перехідні метали
\end{abstract}
\end{document}